\documentclass[%
superscriptaddress,
nofootinbib,
nobibnotes,
amsmath,amssymb,
aps,
floatfix,
]{revtex4-2}

\usepackage{placeins}
\usepackage{booktabs}
\usepackage{dcolumn}
\usepackage{array}
\usepackage{longtable}
\usepackage{float}
\usepackage{hyphenat}
\usepackage{natbib}
\usepackage{hyperref}

\usepackage{amsmath,amsfonts,amsthm,amssymb}
\usepackage{bm,graphicx,graphics,color}
\usepackage{epsf,epsfig}
\usepackage[all]{xy}
\newcommand*{\B}[1]{\ifmmode\bm{#1}\else\textbf{#1}\fi}

\def\bx{\mathbf{x}}

\def\bv{\mathbf{v}}

\def\no{\nonumber}
\def\lb{\label}
\def\be{\begin{equation}}
\def\ee#1{\label{#1}\end{equation}}
\newcommand{\ben}{\begin{eqnarray}}
\newcommand{\een}{\end{eqnarray}}

\usepackage[english]{babel}

\usepackage[letterpaper,top=2cm,bottom=2cm,left=3cm,right=3cm,marginparwidth=1.75cm]{geometry}

\usepackage{amsmath}
\usepackage{graphicx}
\begin{document}

\title{Jeans instability in an expanding universe with dissipation}

\author{Gilberto M. Kremer}
\email{kremer@fisica.ufpr.br}
\affiliation{Departamento de F\'{i}sica, Universidade Federal do Paran\'{a}, Curitiba 81531-980, Brazil}

\begin{abstract}
 Jeans instability is analysed in an expanding universe within the framework of BGK model of the Boltzmann equation and Poisson equations. The background is characterized by a comoving Maxwellian distribution function and a space-time Newtonian gravitational potential which satisfy the BGK model of the Boltzmann and Poisson equations without the necessity to invoke "Jeans swindle".  The perturbations of the distribution function and Newtonian gravitational potentials  from their background states are represented by plane waves of small amplitudes and a differential equation for the density contrast is determined. The density contrast  differential equation was solved numerically and it is shown: (i)  Jeans instability is characterized by  perturbation wavelengths  larger than Jeans wavelength where the density contrast grows with time. The growth of the density contrast is less accentuated for the case where the particle collisions are considered due to an energy dissipation; (ii) for perturbation wavelengths smaller than Jeans wavelength the density contrast has an oscillatory behavior in time and the oscillations for the case where the collisions are taken into account fade away in time due to the energy dissipation.
\end{abstract}

\maketitle

\section{Introduction}

The  determination of the instabilities of self-gravitating fluids from the   hydrodynamic equations of mass and momentum densities coupled with the Newtonian Poisson equation is an old subject in the literature.  It was first analyzed by  Jeans \cite{Jeans} in 1902 who determined a wavelength cutoff from a dispersion relation -- known  nowadays as Jeans wavelength. -- where  the perturbations may propagate as harmonic waves in time for wavelengths  smaller than the Jeans wavelength or may grow or decay in time for  larger  wavelengths. The  Jeans instability refers to the gravitational collapse of  self-gravitating interstellar gas clouds associated with the mass density perturbations which grow exponentially with time \cite{Wein,Coles,BT1}. Physically  a mass density inhomogeneity collapses  when  the outwards pressure force is smaller than the inwards gravitational force.  

 The analysis of Jeans  instability in an expanding Universe by considering  the Newtonian hydrodynamic equations coupled with the Newtonian Poisson equation was due to Bonnor \cite{Bonn} in 1957. This problem is also described in the books \cite{Wein,Coles}.  We call attention to the fact that the Newtonian gravity is still valid in regions whose radius are small compared with the Hubble radius and the velocities  are  non-relativistic.

One can also investigate the Jeans instability  within   the framework of the  collisionless Boltzmann equation coupled with the  Newtonian Poisson equation (see e.g \cite{Coles,BT1,b2,b3,b5,b6,b7,b8,b9}). 

If one considers the collision term in  the Boltzmann equation  irreversible processes  related to  the dissipative effects  of viscosity and heat conductivity  show up. Hence to take into account dissipative effects in the analysis of the Jeans instability one has to consider the collisional Boltzmann equation. Within the framework of a collisional Boltzmann equation Jeans instability was first analyzed in  \cite{b1}. Another methodology to take into account the dissipaitve effects is to consider the hydrodynamic equations for a viscous and heat conducting fluid which was recently analyzed in \cite{krt}.

The aim of the present work is to analyse the Jeans instability from the collisional Boltzmann equation coupled with the  Newtonian Poisson equation in an expanding Universe ruled by a spatially flat Friedmann-Lama\^itre-Robertson-Walker (FLRW) metric. 
Here the BGK model of the Boltzmann equation is consider for the collision term.

 The background distribution function is characterized by a comoving Maxwellian distribution function which satisfies identically the BGK model of the Boltzmann equation. The background Newtonian gravitational potential  satisfies the Poisson equation without the necessity to invoke "Jeans swindle".   The  perturbed one-particle distribution function and  Newtonian gravitational potential from  their  background states are represented by plane waves of small amplitudes. Furthermore, the  perturbed amplitude of the distribution function is consider as a function of the summational invariants \cite{b6,b8,b9}.   A differential equation for the density contrast is obtained from the system of differential equations for the amplitudes. The  numerically solutions of the differential equation  shown that: (i) for perturbation wavelengths  larger than Jeans wavelength the density contrast grows with time which characterizes Jeans instability. The growth of the density contrast is less accentuated for the case where the particle collisions are considered due to an energy dissipation; (ii) for perturbation wavelengths smaller than Jeans wavelength the density contrast has an oscillatory behavior in time. The oscillations for the case where the collisions are taken into account fade away in time due to the energy dissipation.

The paper is structured as follows. In Section \ref{sec2} the the background solution is analyzed.  The Jeans instability is determined in Section \ref{sec3} where the perturbed solution is investigated. The conclusions are given in Section \ref{sec4}.

\section{Analysis of the background solution}\lb{sec2}

The Boltzmann equation describes  the space-time evolution of the one-particle distribution function $f(\bx,\bv,t)$ in the phase space spanned by the  spatial coordinates $\bx$ and velocity $\bv$ of the particles. In the non-relativistic framework it reads (see e.g. \cite{gk})
\ben\lb{1a}
\frac{\partial f}{\partial t}+v_i\frac{\partial f}{\partial x^i}+\frac{\partial f}{\partial v_i}\frac{\partial U}{\partial x^i}=\mathcal{Q}(f,f).
\een
Here $\mathcal{Q}(f,f)$ is the collision operator of the Boltzmann equation which is given by the product of the distribution functions of two particles at binary collisions.

The Boltzmann equation (\ref{1a}) is linked with the Poisson equation for the   Newtonian gravitational potential $U$, namely
\ben\lb{1b}
\nabla^2U=-4\pi G\rho=-4\pi G\int m fd^3v,
\een
where $G$ denotes the universal gravitational constant and $\rho$ the mass density of the fluid. Above the mass density is given in terms of the one-particle distribution function where $m$ denotes  the particle rest mass.

In this work we shall consider the BGK model of the Boltzmann equation where the structure of the collision operator is simplified but preserves the basic properties of the full Boltzmann equation (see e.g. \cite{gk}).  The collision operator in the BGK model  is given in terms of the difference between the one-particle distribution function $f$ and the equilibrium Maxwellian distribution function $f_0$  multiplied by a frequency $\nu$ which is of order of the collision frequency. The  BGK model of the Boltzmann equation  reads 
\ben\lb{1c}
\frac{\partial  f}{\partial t}+v_i\frac{\partial  f}{\partial x^i} +\frac{\partial U}{\partial x^i}\frac{\partial  f}{\partial v^i}=-\nu(f-f_0).
\een

Here we are interested in investigating the Jeans instability in an expanding Universe which is ruled by the spatially flat Friedmann-Lama\^itre-Robertson-Walker (FLRW) metric  $ds^2=(cdt)^2-a(t)(dx^2+dy^2+dz^2)$, where $a(t)$ is the cosmic scale factor. 

In a comoving frame the Maxwellian distribution function is written as
\ben\lb{2a}
f_0(\mathbf{v},t)=\frac{\rho(t)}m\frac1{[2\pi\sigma(t)^2]^{3/2}}\exp\left[-\frac{\left(\mathbf{v}-\dot a\bx_0\right)^2}{2\sigma(t)^2}\right],
\een
by taking into account the Hubble-Lama\^itre's  law ${\bf x}(t)=a(t){\bx_0}$, which relates the comoving coordinates $\bx_0$ with the physical coordinates $\bf x$.  The Maxwellian distribution function  (\ref{2a}) -- which is considered as the background distribution function -- depends on    the fluid mass density $\rho(t)$ and dispersion velocity $\sigma(t)$ which  are functions of time.

The dependence of the background mass density on time is determined from  Friedmann equations, which for a pressureless fluid is given by
\ben\lb{2b}
\rho(t)=\rho_0\left(\frac{a_0}{a(t)}\right)^3,\qquad\hbox{where}\qquad
a(t)=a_0\left(6\pi G\rho_0 t^2\right)^\frac13.
\een

 In terms of the comoving coordinates  the following transformations for  the time and spatial derivatives holds \cite{b8}
\ben\lb{3a}
\frac{\partial}{\partial t}\bigg\vert_{\bx}=\frac{\partial}{\partial t}\bigg\vert_{\bx_0}+\frac{\partial x_0^i}{\partial t}\bigg\vert_{\bx}\frac{\partial}{\partial x^i_0}\bigg\vert_{t}=\frac{\partial}{\partial t}\bigg\vert_{\bx_0}-\frac{\dot a}ax_0^i\frac{\partial}{\partial x^i_0}\bigg\vert_{t},\qquad\frac{\partial }{\partial x^i}\bigg\vert_{t}=\frac1a\frac{\partial }{\partial x_0^i}\bigg\vert_{t},
\een
so that the Boltzmann equation in terms of the comoving coordinates reads
\ben\lb{3b}
\frac{\partial f}{\partial t}\bigg\vert_{\bx_0}+\frac{(v_i-\dot ax_i^0)}a\frac{\partial f}{\partial x_0^i}\bigg\vert_t+\frac1a\frac{\partial U}{\partial x_0^i}\bigg\vert_t\frac{\partial f}{\partial v^i}\bigg\vert_{t,\bx_0}=-\nu(f-f_0).
\een

We assume that the background solution is characterized by the mass density (\ref{2b}) and the Newtonian gravitation potential
\ben\lb{3c}
U_0(\bx,t)=-\frac{2\pi}3G\rho {\bx\cdot\bx}=-\frac{2\pi}3G\rho a^2{\bx_0\cdot\bx_0}.
\een

The insertion of the  background distribution function (\ref{2a}) and gravitational potential (\ref{3c}) into the  Boltzmann equation (\ref{3b}) leads to
\ben\no
&&\frac{\partial f_0}{\partial t}\bigg\vert_{\bx_0}+\frac{(v_i-\dot ax_i^0)}a\frac{\partial f_0}{\partial x_0^i}\bigg\vert_t+\frac1a\frac{\partial U}{\partial x_0^i}\bigg\vert_t\frac{\partial f_0}{\partial v^i}\bigg\vert_{t,\bx_0}=f_0\bigg\{\frac{\dot\rho}\rho+\frac{\left(\mathbf{v}-\dot a\bx_0\right)^2}{\sigma^2}\frac{\dot a}a
\\\lb{3d}
&&+\left[\frac{\left(\mathbf{v}-\dot a\bx_0\right)^2}{\sigma^2}-3\right]\frac{\dot\sigma}\sigma+\frac{\left(\mathbf{v}-\dot a\bx_0\right)\cdot\bx_0}{\sigma^2}\left(\ddot a+\frac{4\pi}3G\rho a\right)\bigg\}=0.
\een
If we take into account the Friedmann equations  for a pressureless fluid and  consider that the dispersion velocity is proportional to the inverse of the cosmic scale factor $\sigma(t)/\sigma_0=a_0/a(t)$, the Boltzmann equation for the background distribution (\ref{3d}) is identically verified.

The Poisson equation (\ref{1b}) is also identically verified for the background  gravitational potential (\ref{3c}), since
\ben\lb{3f}
\nabla^2U_0=-4\pi G\rho=-4\pi G\int mf_0d^3v.
\een
Note that the  background Newtonian  gravitational potential satisfies the Poisson equation  without the necessity to invoke the "Jeans swindle".

\section{Analysis of the perturbed solution}\lb{sec3}

For the determination of the Jeans instability
we require that the  background distribution function (\ref{2a}) and the Newtonian gravitational potential (\ref{3c}) are subjected to small perturbations characterized by $f_1(\bx,\mathbf{v},t)$ and $U_1(\bx,t)$ such that
\ben\lb{4a}
&&f(\bx,\mathbf{v},t)=f_0(\bv,t)+f_1(\mathbf{x},\mathbf{v},t)=f_0(\bv,t)\left[1+h_1(\mathbf{x},\mathbf{v},t)\right],
\qquad
U(\mathbf{x},t)=U_0(\mathbf{x},t)+U_1(\mathbf{x},t).
\een

The perturbations $h_1$ and $U_1$   are represented by plane waves of small amplitudes where the physical wave number vector is ${\bf q}/a(t)$, namely  
\ben\lb{4b}
&&h_1(\mathbf{x},\mathbf{v},t)=\overline h(\mathbf{x},\mathbf{v},t)\exp\bigg(i\frac{\mathbf{q}\cdot\mathbf{x}}{a(t)}\bigg)=\overline h(\mathbf{x},\mathbf{v},t)\exp(i\mathbf{q}\cdot\mathbf{x}_0),
\\\lb{4c}
&&U_1(\mathbf{x},t)=\overline U(t)\exp\bigg(i\frac{\mathbf{q}\cdot\mathbf{x}}{a(t)}\bigg)=\overline U(t)\exp(i\mathbf{q}\cdot\mathbf{x}_0).
\een
The comoving wave number vector is simply $\bf q$ and the factor $1/a(t)$ takes into account that the wavelength is stretched out in an expanding Universe. The above amplitudes $\overline{h}(\bx,\bv,t)$ and $\overline {U}(t)$ are considered to be small. 

In the kinetic theory of gases an important quantity is the so-called summational invariant which is conserved in a binary encounter of the particles. In the non-relativistic framework the summational invariants are: the rest mass $m$, the momentum $m\bv$ and the energy $mv^2/2$ of a particle. We follow \cite{b6} and assume that $\overline h$ is given as a linear combination of the comoving summational invariants  1, $\left(\mathbf{v}-\dot a\bx_0\right)$ and $\left(\mathbf{v}-\dot a\bx_0\right)^2$, namely
\ben\lb{4d}
\overline h(\mathbf{x},\mathbf{v},t)=A(t)+\mathbf{B}(t)\cdot\left(\mathbf{v}-\dot a\bx_0\right)+D(t)\left(\mathbf{v}-\dot a\bx_0\right)^2.
\een
The unknown functions of time  $A(t)$, ${\bf B}(t)$ and $D(t)$ do not depend on the comoving summational invariants.

We insert (\ref{4a}) -- (\ref{4d})  into the  Boltzmann (\ref{3b}) and Poisson equations (\ref{1b}) and get the following system of equations:
\ben\no
&&f_0\left[\frac{\partial \overline h}{\partial t}\bigg\vert_{\bx_0}+\frac{\left(v_i-\dot ax_0^i\right)}a \frac{\partial \overline h}{\partial x_0^i}\bigg\vert_{t}+\frac1a\frac{\partial U_0}{\partial x_0^i}\bigg\vert_{t}\frac{\partial\overline h}{\partial v^i}\bigg\vert_{t,\bx_0}\right]+\frac1a\frac{\partial U_1}{\partial x_0^i}\bigg\vert_{t}\frac{\partial f_0}{\partial v^i}\bigg\vert_{t,\bx_0}=f_0\bigg\{\frac{dA}{dt}+\left(\mathbf{v}-\dot a\bx_0\right)\cdot\frac{d{\bf B}}{dt}
\\\no
&&\qquad+\left(\mathbf{v}-\dot a\bx_0\right)^2\frac{dD}{dt}
-\frac{\dot a}a\left(\mathbf{v}-\dot a\bx_0\right)\cdot\left[{\bf B}+2D\left(\mathbf{v}-\dot a\bx_0\right)\right]+\frac{i{\bf q}\cdot\left(\mathbf{v}-\dot a\bx_0\right)}{a}\bigg[A+\mathbf{B}\cdot\left(\mathbf{v}-\dot a\bx_0\right)
\\\no
&&\qquad+D\left(\mathbf{v}-\dot a\bx_0\right)^2-\frac{\overline U}{\sigma^2}\bigg]
-\underline{\left(\ddot a+\frac{4\pi}3\rho a\right)}\bx_0\cdot\left[{\bf B}+2D\left(\mathbf{v}-\dot a\bx_0\right)\right]\bigg\}
\\\lb{5a}
&&\qquad=-\nu\left[A+\mathbf{B}\cdot\left(\mathbf{v}-\dot a\bx_0\right)+D\left(\mathbf{v}-\dot a\bx_0\right)^2\right],
\\\lb{5b}
&&\frac{q^2}{a^2}\overline U=4\pi G\int m f_0\bigg[A+\mathbf{B}\cdot\left(\mathbf{v}-\dot a\bx_0\right)
+D\cdot\left(\mathbf{v}-\dot a\bx_0\right)^2\bigg]d^3v= 4\pi G \rho\left(A+3\sigma^2D\right).
\een
 The underlined term in (\ref{5a}) vanishes since it refers to the expression of acceleration equation for a pressureless fluid in the Friedmann equations.  The last equality in (\ref{5b}) was determined through integration by using   Gaussian integrals.

 From the successive multiplication of (\ref{5a}) by the comovig summational invariants  1, $(\mathbf{v}-\dot a\bx_0)$ and $\left(\mathbf{v}-\dot a\bx_0\right)^2$ and integration of the resulting equations by using  Gaussian integrals  follows the system of differential equations 
\ben\lb{6a}
&&\frac{dA}{dt}+3\sigma^2\frac{dD}{dt}+i\frac{\sigma^2}{a}B-6\frac{\dot a}a\sigma^2D=-\nu\left(A+3\sigma^2D\right),
\\\lb{6b}
&&\frac{dB}{dt}+i\frac{q^2}a\left[A+5\sigma^2D-\frac{\overline U}{\sigma^2}\right]-\frac{\dot a}aB=-\nu B,
\\\lb{6c}
&&3\frac{dA}{dt}+15\sigma^2\frac{dD}{dt}+i5\frac{\sigma^2}{a}B-30\frac{\dot a}a\sigma^2D=-\nu\left(3A+15\sigma^2D\right).
\een
Note that  in the above equations we have introduced $B(t)=\mathbf{B}(t)\cdot \mathbf{q}$ and that (\ref{6b}) results from the  scalar multiplication of the integrated equation by $\bf q$.

Let us analyse the system of differential equations (\ref{6a}) -- (\ref{6c}). First the subtraction  of  (\ref{6c}) from (\ref{6a}) multiplied by five results that $dA/dt=-\nu A$ which implies that  $A=e^{-\nu t}$.

Next we introduce the density contrast defined by
\ben\lb{7a}
\delta\rho=\frac{\int mf_0\overline h d^3v}{\rho}=A+3\sigma^2D,
\een
so that  (\ref{6a}) can be rewritten in terms of the density contrast as
\ben\lb{7b}
\frac{d\delta\rho}{dt}+i\frac{\sigma^2}aB=-\nu\delta\rho.
\een
Above the relationship  $\sigma/\sigma_0=a_0/a$ was considered.

From the differentiation of (\ref{7b}) with respect to time and the elimination of $B$, $dB/dt$ and $\overline U$ by using  (\ref{6a}),  (\ref{6b})  and (\ref{5b}), respectively,  the following differential equation for the density contrast is obtained
\ben\lb{8a}
\frac{d^2\delta\rho}{dt^2}+\bigg(2\frac{\dot a}a+\nu\bigg)\bigg(\frac{d\delta\rho}{dt}+\nu\delta\rho\bigg)+\bigg(\frac{5q^2\sigma^2}{3a^2}-4\pi G\rho\bigg)\delta\rho{-\frac{2e^{-\nu t}q^2\sigma^2}{3a^2}}=-\nu\frac{d\delta\rho}{dt}.
\een

Now we introduce the dimensionless quantities 
\ben\lb{8b}
\lambda_J=\frac{2\pi \sqrt{5/3}\,\sigma}{\sqrt{4\pi G\rho}},\qquad \lambda=\frac{2\pi a_0}q,\qquad \tau=t\sqrt{6\pi G \rho},
\een
where $\lambda_J$ is the Jeans wavelength, $\lambda$ is related with the wavelength of the perturbation and $\tau$ is a dimensionless time.

In terms of the dimensionless quantities (\ref{8b}) the differential equation for the density contrast  (\ref{8a}) becomes 
\ben\lb{9}
\delta\rho''+\bigg(\frac4{3\tau}+2\nu_*\bigg)\delta\rho'+\bigg[\nu_*\bigg(\frac4{3\tau}+\nu_*\bigg)+\frac23\bigg(\frac{\lambda_J^2}{\lambda^2\tau^\frac43}-1\bigg)\bigg]\delta\rho-\frac{4\lambda_J^2}{15\lambda^2}\frac{e^{-\nu_*\tau}}{\tau^\frac43}=0.
\een
Above the prime refers to the differentiation  with respect to $\tau$,  moreover the dimensionless collision frequency  $\nu_*=\nu/\sqrt{6\pi G\rho}$ was introduced and 
the relationships $a'/a=2/3\tau$ and $a=a_0\tau^\frac23$ were used.

\begin{figure}[ht]
\centerline{\includegraphics[width=12cm]{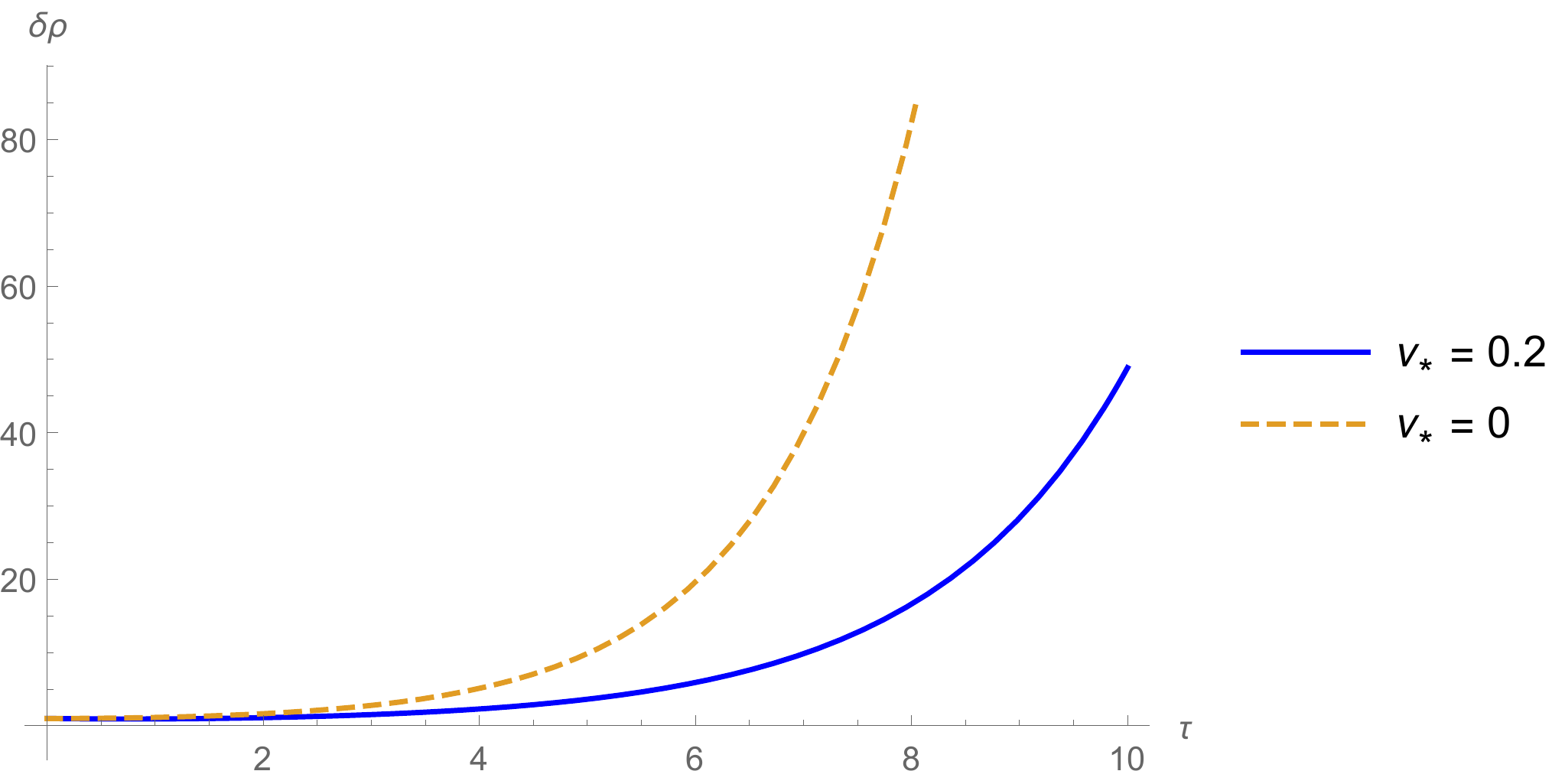}}
\caption{Density contrast  $\delta\rho$ as function of the dimensionless time $\tau$  when  $\lambda/\lambda_J=10$, for the dimensionless collision frequencies $\nu_*=0.1$ and $\nu_*=0$.}\lb{fig1}
\end{figure}

The differential equation (\ref{9}) was solved numerically with the initial conditions $\delta\rho(10^{-3})=1$ and $\delta\rho'(10^{-3})=0$. The solutions of the differential equation  for the  density contrast $\delta\rho$ as function of the dimensionless time $\tau$ are plotted in Figs. \ref{fig1}  and \ref{fig2}.  The straight lines  represent the case $\nu_*\neq0$ where the collisions are taken into account while the dashed lines correspond to a collisionless Boltzmann equation where $\nu_*=0$.

In Fig. \ref{fig1} the ratio between the Jeans wavelength $\lambda_J$ and the one associated with the perturbation $\lambda$ was chosen as $\lambda/\lambda_J=10$, the straight line corresponds to $\nu_*=0.2$ while the dashed line to $\nu_*=0$ (collisionless Boltzmann equation). In this case the perturbation wavelength is bigger than the Jeans wavelength and the density contrast grows with time which corresponds to Jeans instability. We note that due to the presence of the particle  collisions an energy dissipation comes out implying  a less accentuate growth of the density contrast in comparison to the one for a collisionless Boltzmann equation represented by  $\nu_*=0$.

For the case where  the perturbation wavelength is smaller than the Jeans wavelength, the density contrast has an oscillatory behavior in time. In Fig. \ref{fig2} it is shown the time evolution of the density contrast for $\lambda/\lambda_J=0.1$, the straight line corresponding to $\nu_*=0.5$ and the dashed one to $\nu_*=0$. We note that the oscillatory behavior of density contrast is damped for the case where the particle collisions are taken into account due to the energy dissipation and in this case the density contrast fades away for big times.

\begin{figure}[ht]
\centerline{\includegraphics[width=12cm]{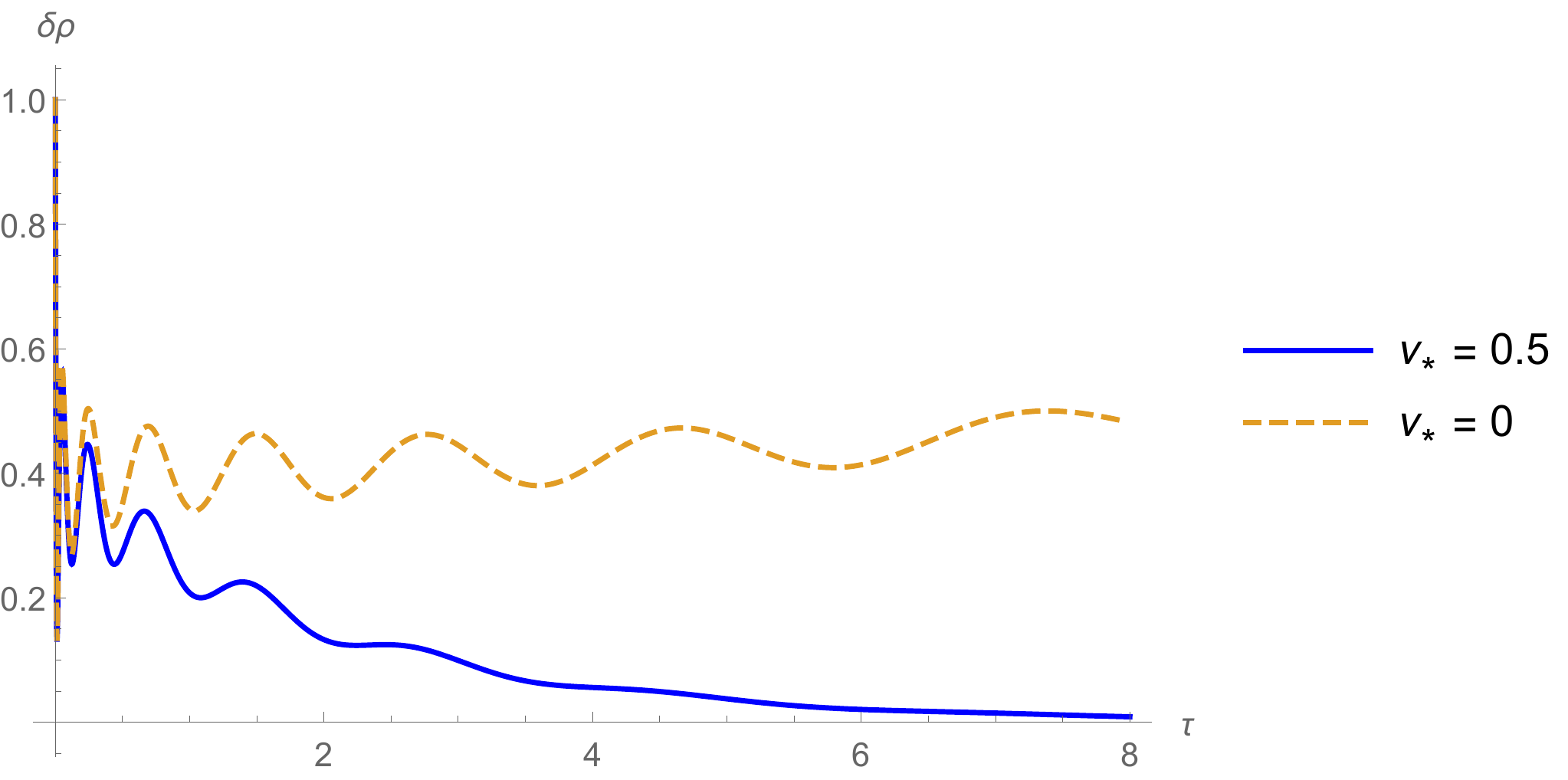}}
\caption{Density contrast  $\delta\rho$ as function of the dimensionless time $\tau$  when  $\lambda/\lambda_J=0.1$, for the  dimensionless collision frequencies $\nu_*=0.1$ and $\nu_*=0$.}\lb{fig2}
\end{figure}

\section{Conclusions}\lb{sec4}
In this work the Jeans instability in an expanding Universe ruled by a spatially flat FLRW metric was analyzed within the framework of the BGK model of Boltzmann equation -- where the collisions of the particles were taken into account -- and Newtonian Poisson equation. The background distribution function -- characterized by a comoving Maxwellian distribution -- satisfies identically the BGK model of the Boltzmann equation. There is no necessity to invoke "Jeans swindle", since the background Newtonian gravitational potential  satisfies the Poisson equation.   The  one-particle distribution function and the Newtonian gravitational potential were perturbed from  their  background states represented by plane waves of small amplitudes. The perturbed amplitude of the distribution function was considered as a function of the summational invariants. From the system of differential equations for the amplitudes  a differential equation for the density contrast was obtained. This differential equation was solved numerically and it was shown that: (i) for perturbation wavelengths  larger than Jeans wavelength the density contrast grows with time which characterizes Jeans instability. The growth of the density contrast is less accentuated for the case where the particle collisions are considered due to an energy dissipation; (ii) for perturbation wavelengths smaller than Jeans wavelength the density contrast has an oscillatory behavior in time. The oscillations for the case where the collisions are taken into account fade away in time due to the energy dissipation.

\acknowledgments{ This work was supported by Conselho Nacional de Desenvolvimento Cient\'{i}fico e Tecnol\'{o}gico (CNPq), grant No.  304054/2019-4.}

\end{document}